\documentclass[a4paper,fleqn,usenatbib]{mnras}
\usepackage{newtxtext,newtxmath}
\usepackage[T1]{fontenc}
\usepackage{ae,aecompl}

\usepackage{amsmath}
\usepackage[final]{graphicx}
\usepackage{amssymb}

\newcommand{\real}{\mathbb R}
\DeclareMathOperator{\occ}{occ}

\DeclareMathOperator{\divergence}{div}
\DeclareMathOperator{\Kn}{Kn}
\newcommand{\aprilXV}{April 2015}
\newcommand{\juneXV}{June 2015}
\newcommand{\septemberXV}{September 2015}
\newcommand{\mayXVI}{May 2016}

\title[Seasonal changes of the volatile density of 67P]{Seasonal changes of the volatile density in the coma and on the surface of comet 67P/Churyumov-Gerasimenko}
\author[T. Kramer et al.]{
Tobias Kramer,$^{1,2}$\thanks{E-mail: kramer@zib.de}
Matthias L\"auter,$^{1}$
Martin Rubin$,^{3}$
Kathrin Altwegg$^{3}$
\\
$^{1}$Zuse Institute Berlin, 14195 Berlin, Germany\\
$^{2}$Department of Physics, Harvard University, Cambridge, MA, USA\\
$^{3}$University of Bern, Space Research and Planetary Sciences, 3012 Bern, Switzerland
}

\date{Accepted 2017 April 5. Received 2017 April 3; in original form 2017 February 17.}

\pubyear{2017}

\begin{document}
\label{firstpage}
\pagerange{\pageref{firstpage}--\pageref{lastpage}}
\maketitle

\begin{abstract}
Starting from several monthly data sets of Rosetta's COmetary Pressure
Sensor we reconstruct the gas density in the coma around
comet 67P/Churyumov-Gerasimenko.
The underlying inverse gas model is constructed by fitting ten
thousands of measurements to thousands of potential gas sources
distributed across the entire nucleus surface.
The ensuing self-consistent solution for the entire coma density and
surface activity reproduces the temporal and spatial variations seen
in the data for monthly periods with Pearson correlation coefficients
of $0.93$ and higher.
For different seasonal illumination conditions before and after
perihelion we observe a systematic shift of gas sources on the nucleus.
\end{abstract}


\section{Introduction}

The Rosetta mission provides a systematic study of the comet
67P/Churyumov-Gerasimenko (67P/C-G) on its way across the inner solar
system.
In contrast to previous fly-by missions, Rosetta monitored the
cometary environment constantly with a wide variety of instruments
\citep{Schulz2009}.
The nucleus of comets consists of a mixture of dust and ice \citep{Huebner2006}.
Upon approaching the inner solar system, the nucleus heats up,
frozen volatiles sublimate, and gas together with dust is released.
The continuous emission results in a coma, consisting of neutral and ionised species, and gives rise to a tail region surrounding the nucleus.
Depending on the composition, size, and activity of the nucleus, the gas and dust densities vary considerably.
For 67P/C-G approaching the sun, the total gas production rate
increased with peak activity occurring $\sim 20$ days after perihelion
\citep{Hansen2016}.
The COmetary Pressure Sensor (COPS) on-board of Rosetta has probed the
gas density and dynamic pressure over a period of several months and thus
provides a significant data set for a comet coma with a high
temporal resolution in 1~minute intervals \citep{Balsiger2007}.
Due to the spreading and mixing of gas in space and the unknown surface emission rates 
it is not possible to deduce from the in-situ coma measurements directly to the local
surface activity at the sub-spacecraft points.
The surface emission from different points on the nucleus is the input to any model of the entire coma \citep{Combi2004}.
The deduction of surface properties from coma features requires to solve the inverse problem,
which is formulated here as an optimisation procedure.
The complexity of the inversion is increased by the need of an accurate gas model and by the large amount of in-situ data points.
For instance, the direct simulations Monte Carlo (DSMC) method brings along most accurate atomistic results \citep{Bird1994}, but cannot be applied to study all possible surface emission scenarios to be compared with the COPS data \citep{Fougere2016}.
In addition, the complicated non-convex nucleus shape of 67P/C-G
does not admit model reductions by symmetry considerations and further increases 
the computational complexity.
Here, we present a different approach to describe the coma, which allows us to 
assimilate the large amount of measurements collected by COPS on Rosetta 
at moderate computational costs.
We carry out the inversion and optimisation procedure to find the converged, 
self-consistent gas field and the corresponding distribution of 
surface emitters on 67P/C-G.
To stay numerically efficient, we work with simplified gas equations and neglect collisions in the tenuous atmosphere around 67P/C-G.
The converged source distribution minimises the least squares deviation of the measured vs.\ modeled density at Rosetta's position and reproduces the COPS data set over several weeks with high Pearson correlation coefficients of $0.93$--$0.99$.\\
The articles is organised as follows:
In Sect.~\ref{sec:model} we describe the mathematical model and the
optimisation procedure and in Sect.~\ref{sec:results} we discuss the
resulting gas emission from surface sources.
We apply the model to different heliocentric distances of 67P/C-G and discuss
the ensuing gas field for varying sub-solar latitudes, corresponding to
pre-equinox (\aprilXV{}), equinox (\juneXV{}),
shortly after perihelion (\septemberXV{}), and 9 months
past-perihel (\mayXVI{}) conditions.
Finally, we describe how the gas sources are shifting with the changing
seasonal solar illumination along the orbit of 67P/C-G and how the sources correlate with optical Rosetta observations.

\section{Cometary gas models}\label{sec:model}

A common ancestor of all cometary gas models is the \cite{Haser1957} model for gas emitted constantly in time from a
spherical nucleus and with a constant radial velocity.
Neglecting ionization, the equation of continuity implies a $x^{-2}$ drop of the
neutral gas density with distance $x$ from the comet.
Other coma models consider a more complex gas dynamics around a non-spherical nuclear shape.
Here, we are considering the inner coma up to a distance of 400~km.
Up to this distance, the gas distributes within $\sim 400$~s and retardation effects due to the rotation of the nucleus can be neglected.
The gas density in this space increases from $10^{15}$~molecules/m$^{3}$ in \aprilXV{} (heliocentric distance $1.8$~AU) to $10^{18}$~molecules/m$^{3}$ two weeks after perihelion (heliocentric distance $1.3$~AU).
For the months excluding perihelion, this implies a relatively large mean free path length 
$l_{\rm mfp}=5000$--$50$~m and Knudsen number $\Kn=l_{\rm mfp}/L=0.1$--$10$
and indicates that collisions occur less frequently compared to more active comets, 
where ${l'}_{\rm mfp}  = 0.1$--$1$~m \citep{Gombosi1986}.
The small amount of collisions allows us to build up a coma model from the analytical gas equations for the collisionless case \citep{Narasimha1962}.
The analytical gas solutions match the DSMC results for high Knudsen numbers $\Kn=l_{\rm mfp}/L \gg 1$
\citep{Cai2012}, while at lower Knudsen numbers $\Kn\sim 0.1$ the error is below 19\%.
For the highest gas densities encountered in the weeks after perihelion, the
collisionless approximation introduces a larger relative error of $0.18$, but
still fits the observations with a Pearson correlation $0.93$, see
Tab.~\ref{tab:fitparam}.
\begin{figure}
\begin{center}
\includegraphics[width=0.8\columnwidth]{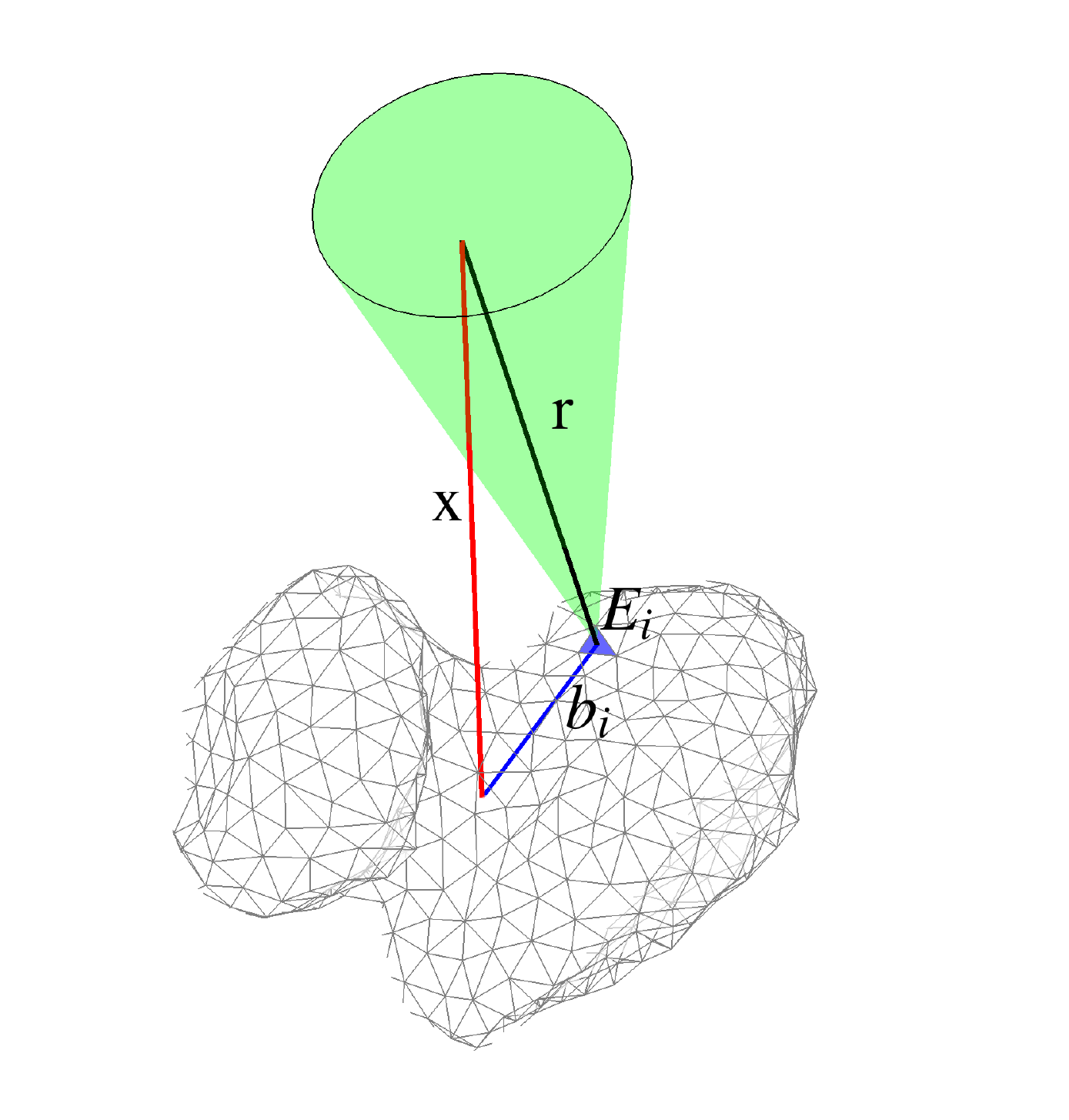}\\
\end{center}
\caption{Low resolution triangular shape approximation $S^{(1024)}$ for the comet 67P/C-G and coordinate system for an exemplary gas source located at the highlighted triangular face $E_i$.
}\label{fig:cone}
\end{figure}

\subsection{Contribution of a single surface face}\label{ssec:single}

For the presented model, the shape of 67P/C-G with concave areas is
approximated by a high resolution triangular surface mesh.
Each mesh face acts a potential gas source, albeit with a position
dependent activity, Fig.~\ref{fig:cone}.
The momentary gas density at the spacecraft position $\mathbf{x}\in\real^3$ results from the
linear superposition of all visible gas sources, possibly located on different parts of the bi-lobed nucleus.
As a consequence, the dominant factor for the variability in the gas
density around 67P/C-G is the irregular shape of the nucleus.
Let us assume a piecewise triangular shape approximation $S^{(n)}$ for the
surface based on $n$ triangular surface elements $E_i$, namely
\[
S^{(n)} = \bigcup_{i=1}^n E_i.
\]
Each face $E_i$ emits collisionless gas into a surrounding vacuum
with a point gas-source centered at
$\mathbf{b}_i=\text{center of mass}(E_i)$ and pointing along the outward surface normal
$\boldsymbol{\nu}_i$.
For this case, \cite{Narasimha1962} derives the analytical results for a gas temperature $T$ and outflow velocity $u_0$.
Introducing the parameter $\beta = 1 / (2 R T)$,  
with specific gas constant $R$ and $U_0=u_0 \sqrt{\beta}$,
the density profile at a space point 
$\mathbf{x}=\mathbf{b}_i+\mathbf{r}$ is given by
\begin{equation}\label{eq:gasfield}
\rho_i(\mathbf{x})=
\begin{cases}
\frac{U_0}{\pi}\frac{\cos\theta}{r^2}
|E_i|q_i\exp(-U_0^2\sin^2\theta), & \text{if~} \cos\theta>0\\
0, & \text{otherwise}
\end{cases}.
\end{equation}
Here, $r=|\mathbf{r}|$, $\cos\theta = \mathbf{r}\cdot \boldsymbol{\nu}_i / r$
and the source strength is denoted by $q_i$.
With the corresponding velocity field
\begin{equation}\label{eq:velfield}
\mathbf{u}_i(\mathbf{x})=\frac{\mathbf{r}}{r}\frac{U_0}{\sqrt{\beta}}\cos(\theta),
\quad\text{for}\quad\cos\theta>0,
\end{equation}
the momentum field becomes free of sources, that is
$\divergence( \rho_i \mathbf{u}_i ) = 0$.
The momentum field carries along the molecular production rate
$\dot{\rho}_i$ from a source on face $E_i$ at any distance $\epsilon>0$
\begin{equation}\label{eq:cont}
\dot{\rho}_i =
\int_{\mathbf{r}\cdot\boldsymbol{\nu}_i=\epsilon}
\rho_i \mathbf{u}_i \cdot \boldsymbol{\nu}_i  {\rm d} \sigma
= \left(1- \frac{F(U_0)}{U_0}\right) \frac{|E_i| q_i}{\sqrt{\beta}}.
\end{equation}
The factor $1-F(U_0)/U_0$, with Dawson's integral $F$, approaches 1
for $1< U_0$.
In this limit $|E_i|q_i/\sqrt{\beta}$ is the production rate
$\dot{\rho}_i$, and
\begin{equation}\label{eq:faceprod}
\dot{\rho}_{i,{\rm surf}} = \frac{q_i}{\sqrt{\beta}}
\end{equation}
becomes the surface emission rate (flux) with the unit molecules per
time and surface area.
In general, the gas variables $\rho_i$ and $\mathbf{u}_i$
are functions in space and time and depend on the previously and
momentary received solar radiation.
We consider the time-averaged (over several weeks) 
surface emission rate and  do not consider oscillations caused by the 
momentary illumination conditions.
This restriction clearly brings out the spatial variations
of the surface emission rate and
does not introduce a large error in the description
of each monthly COPS data set, as will be shown below.
\begin{figure*}
\begin{center}
\includegraphics[width=0.9\textwidth]{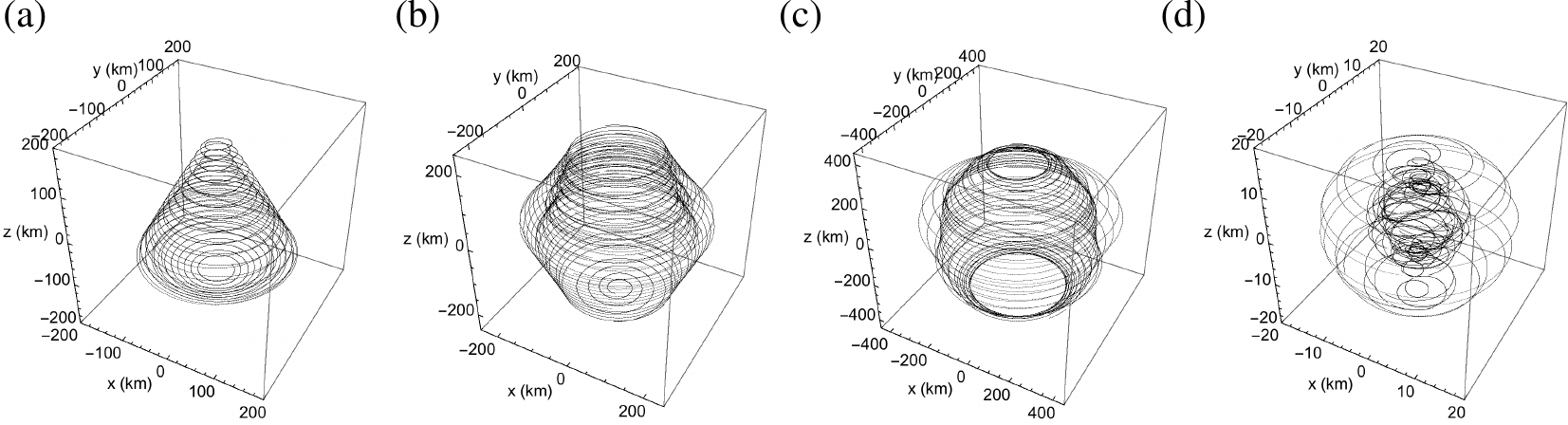}
\end{center}
\caption{Rosetta's orbit in the comet-fixed frame for the four COPS data sets considered. 
(a) April 12-31 2015, (b) \juneXV{}, (c) \septemberXV{}, (d) \mayXVI{}.
In all cases Rosetta surveys the entire surface of the nucleus, which rotates with a period of $\sim 12.4$~h.
}
\label{fig:orbit}
\end{figure*}

\subsection{Construction of the entire gas coma}

The gas emission of the complete shape $S^{(n)}$ is described by the linear superposition of all face contributions \eqref{eq:gasfield}.
The model builds on an accurate shape representation of the nucleus, given by a regular triangular surface mesh and including concave areas.
We use a shape model with 9996 nearly equally sized triangles derived by uniform remeshing from the high resolution model constructed by  \cite{MalmerShapeModelNovmeber2015}.
The bi-lobed shape and the concavities found on the surface of 67P/C-G require to consider the obstruction of gas sources according to the spacecraft location.
Since the gas expands along straight lines, 
Eq.~\eqref{eq:velfield}, a face $E_i$ only contributes to an observation point $\mathbf{x}\in\real^3$ if the line between $\mathbf{b}_i$ and $\mathbf{x}$ does not cross
the surface of the shape $S^{(n)}$.
Summing up over all faces gives the gas density around the comet:
\begin{equation}\label{eq:shapesource}
\rho(\mathbf{x}) = \sum_{i=1}^n \occ_i(\mathbf{x})\rho_i(\mathbf{x}),\quad
\rho \mathbf{u}(\mathbf{x})
= \sum_{i=1}^n \occ_i(\mathbf{x})\rho_i \mathbf{u}_i (\mathbf{x}),
\end{equation}
\begin{equation*}
\occ_i(\mathbf{x}) =
\begin{cases}
0, & \text{if the line between $\mathbf{b}_i$ and $\mathbf{x}$ crosses $S^{(n)}$} \\
1, & \text{otherwise}
\end{cases}.
\end{equation*}
Neglecting any molecular desorption or adsorption, the total
production rate $Q$ of the the comet shape $S^{(n)}$ arises from
the sum of all source contributions.
The outflow across a reference sphere of radius $R>0$ and with surface
element ${\rm d}\sigma$ is given by
\begin{equation}\label{eq:total}
Q = \int_{|\mathbf{x}|=R} \rho\mathbf{u}
\cdot \boldsymbol{\nu} \; {\rm d} \sigma.
\end{equation}
For the simplified case of a convex comet shape ($\occ_i=1$) we obtain
\begin{equation*}
Q_{\rm convex} = \sum_{i=1}^n \dot{\rho}_i.
\end{equation*}
If we assume in addition a spatially homogeneous activity $\dot{\rho}_{i} = Q_0 / n$
for all faces $i=1,...,n$, one recovers the \cite{Haser1957} model
with total production rate given by
\begin{equation*}
Q_{\rm Haser} = 4 \pi R^2 u_0 \rho(R)
= Q_0.
\end{equation*}
For 67P/C-G the last two simplifications do not seem appropriate and
we use the general expression~(\ref{eq:total}).

\begin{table}
\centering
\begin{tabular}{l|l|l|l}
           & Rosetta
 & sub-solar  & heliocent. \\ 
           & dist. [km]
 & latitude & dist. [AU] \\ \hline\hline
\aprilXV{} & $179$--$91$ &
$10.4^\circ$--$4.1^\circ$
 & $1.89$--$1.75$ \\\hline
equinox\\ May~11~2015 & & $0^\circ$ & \\\hline
\juneXV{} & $160$--$234$ & $-9.5^\circ$-- $-25.3^\circ$
& $1.52$--$1.35$ \\\hline
perihelion\\ Aug~13~2015 & & $-47.8^\circ$ & $1.24$ \\\hline
Aug~22--Sep~18~2015 & $312$--$442$ & $-50.8^\circ$-- $-52.3^\circ$
& $1.25$--$1.32$ \\\hline
summer solstice\\
Sep~5~2015 &  & $-52.3^\circ$ & $1.28$ \\\hline
equinox\\ March~23~2016 & & $0^\circ$  & \\\hline
\mayXVI{} & $7.1$--$18.6$ & $5.8^\circ$--$9.6^\circ$ &
$2.91$--$3.12$
\end{tabular}
\caption{Summary of the distance Rosetta--comet, sub-solar latitude, and heliocentric distance for all analysed months.}
\label{ta:months}
\end{table}

\begin{table}
\centering
\begin{tabular}{l|rrrr}
monthly            & April   & June    & September & May \\ 
dataset            & 2015    & 2015    & 2015 & 2016 \\\hline
COPS samples ($N$) & $21125$ & $32860$ & $32710$ & $24817$ \\
gas species        & H$_2$O  & H$_2$O  & H$_2$O  & CO$_2$ \\
Pearson correlation& $0.97$  & $0.93$  & $0.93$  & $0.99$ \\
err from Eq.~\eqref{eq:relerror} & $0.12$ & $0.18$  & $0.18$  & $0.11$ \\
\end{tabular}
\caption{Parameters for the gas model coupled to $n=9996$ triangular faces to represent the shape of comet 67P/C-G.}\label{tab:fitparam}
\end{table}

\subsection{Parameter evaluation for the inverse gas model}\label{ssec:fit}

The gas distribution in the coma depends on the parameters
$U_0$ and $q_i$ in \eqref{eq:gasfield}, to be specified for all $i=1,...,n$ faces.
The inverse problem consists of determining these $n+1$
parameters by making use of the larger set of $N$ measured gas densities $\rho_j$ in the coma at different spatial positions $\mathbf{x}_j$.
\begin{equation*}
(\mathbf{x}_j,\rho_j) \in \real^3\times\real,\quad \text{for}\quad j=1,...,N.
\end{equation*}
For $n+1<N$, the system of equations relating the $n$ faces to the $N$
observations $\boldsymbol{\rho}=(\rho_1,...,\rho_N)$ results in an
overdetermined problem and is nonlinear in $U_0$.
Applying Eq.~\eqref{eq:shapesource}, we abbreviate this parametric
dependence by
\begin{equation*}\label{eq:hits}
\rho_{U_0,\mathbf{q}}(\mathbf{x}_i) = \rho_i,\quad i=1,...,N
\end{equation*}
with respect to $U_0$ and the unknown surface activity parameters
$\mathbf{q}=(q_1,...,q_n)$.
The problem statement is readily expressed in matrix form by
$M_{U_0}\mathbf{q} = \boldsymbol{\rho}$,
with the model matrix $M_{U_0}\in\real^{n\times N}$
containing the contribution of every surface triangle to a given
COPS location.
We require $(U_0,\mathbf{q})\in\real^{n+1}$ to give the minimal error between observation data and model 
\begin{equation*}
\min_{({U_0}',\mathbf{q}')\in\real^{n+1}}
\|M_{{U_0}'} \mathbf{q}'-\boldsymbol{\rho}\|_{l^2(\real^N)}.
\end{equation*}
Having found this optimal solution set $U_0$ and $\mathbf{q}$, the gas field results from Eq.~\eqref{eq:shapesource}.
The relative $l^2$-error for the optimal solution is given by
\begin{equation}\label{eq:relerror}
err =
\frac{\|M_{U_0} \mathbf{q}-\boldsymbol{\rho}\|_{l^2(\real^N)}}
{\|\boldsymbol{\rho}\|_{l^2(\real^N)}}.
\end{equation}
The same inverse optimisation procedure could be applied to other gas
models.
For the analytical gas model used here, Eq.~(\ref{eq:gasfield}), 
the matrix $M_{U_0}$ is constructed very efficiently.

\section{Observational data and model results}\label{sec:results}

The model is evaluated directly at Rosetta's sampling points and at a reference sphere with radius 100~km. 
Finally, we give the
corresponding surface emission rate on the cometary surface.
As input we take the observational data of Rosetta's COPS
for four months in 2015 and 2016.
COPS consists of two gauges. The Nude Gauge measures the total neutral gas density and the Ram Gauge records the dynamic pressure of the gas that emanates from the nucleus \citep{Balsiger2007}.
For this work, COPS Nude gauge measurements are used.
In the Nude Gauge a hot filament emits electrons that are then accelerated towards the anode grid in which they ionise neutral particles.
The resulting ions are measured  with a sensitive electrometer.
For the data set shown here, first the spacecraft background gas surrounding Rosetta \citep{Schlaeppi2010} is subtracted to obtain the density of the cometary volatiles.
COPS calibration in the laboratory has been performed with molecular nitrogen N$_2$.
The measured densities at the comet have been corrected for different compositions to account for the species dependent instrument sensitivities with respect to N$_2$.
Around 67P/C-G, water vapor H$_2$O is the most abundant species  
besides CO and CO$_2$ at larger heliocentric distance \citep{Fougere2016}.
For simplicity, we assume a one component coma, which introduces errors on the order of 10\%.
On top of that, the errors in COPS absolute densities are approximately 15\%, yielding a total error of 18\%.

\begin{figure*}
\begin{center}
\includegraphics[width=0.75\textwidth]{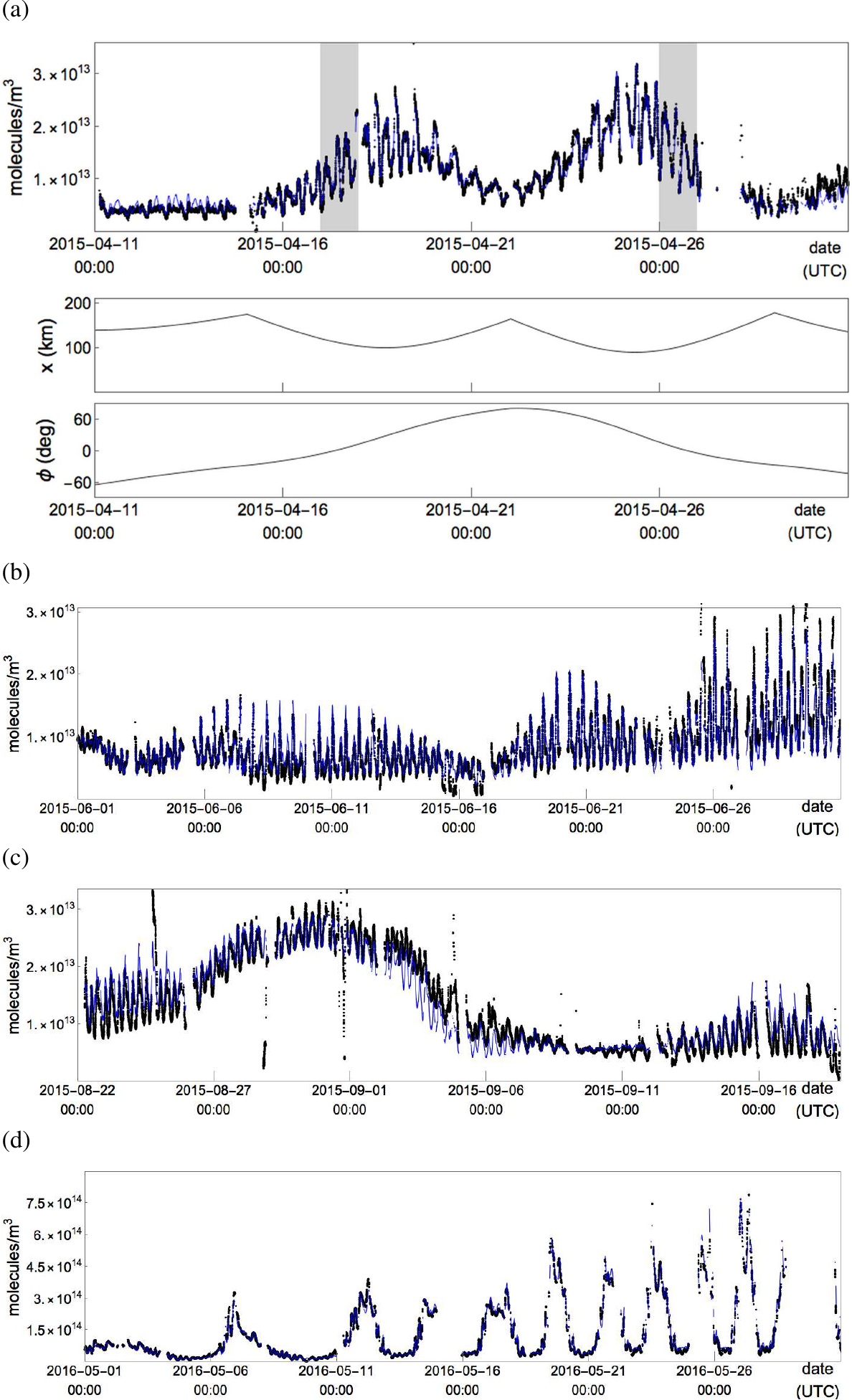}
\end{center}
\caption{
Modeled gas number density (blue lines) vs.\ COPS data (black dots) for (a) \aprilXV{}, (b) \juneXV{}, (c) \septemberXV{}, and (d) \mayXVI{}. 
Narrow spikes (instrument pointing and thruster operation related) in the COPS data have been removed.
(a) The shaded time intervals are shown in detail in Fig.~\ref{fig:copshourly}.
The sub panels display the distance $x$ of Rosetta from the nucleus
and the sub-spacecraft latitude $\phi$.
}
\label{fig:copsfit}
\end{figure*}

\subsection{Inverse gas model for COPS data}\label{ssec:inverse}

We consider data sets for four distinct months, 
\aprilXV{}, \juneXV{}, \septemberXV{}, and \mayXVI{}.
Each of the data set represents a characteristic phase of the
cometary orbit with respect to sub-solar latitude,
see Tab.~\ref{ta:months}.
The inbound equinox occurs between the first two considered months, April and June 2015, 
which are still before perihelion.
The \septemberXV{} set covers the highest activity period directly after perihelion.
The last COPS data set from \mayXVI{} surveys a period after
outbound equinox and 9 months after perihelion.
For all observation months,
Rosetta surveyed the entire nucleus with all surface areas
within sight, Fig.~\ref{fig:orbit}.
For April and June 2015, the distance from
Rosetta to 67P/C-G is above 100~km, in \septemberXV{} it increases to 400~km, 
and for \mayXVI{} the distance stays below 20~km.
All model results and least-square fits are performed for the shape model $S^{(n)}$ with $n=9996$
and $N$ COPS density measurements, summarised in Tab.~\ref{tab:fitparam}. 
For all cases, the relative error Eq.~\eqref{eq:relerror} is in the
range $0.11$--$0.18$, which results in high Pearson correlation
coefficients ($0.93$--$0.99$) between the measured data and the model.
The high correlation coefficients show that the ansatz of a linear superposition of gas sources provides an accurate parametrisation of the COPS data set, 
despite the assumption of a stationary and collisionless gas flow.
The remaining differences between the inverse model and the
observations stem from sporadic outbursts in gas emissions,
instrumental uncertainties, and gas collisions.
In Fig.~\ref{fig:copsfit}, we show a direct comparison of the measured
number density and the model results from Eq.~(\ref{eq:shapesource})
for (a) \aprilXV{}, (b) \juneXV{}, (c) \septemberXV{}, and (d) \mayXVI{}.
The overall increase of the gas density around Apr 16-21 and Apr 23-29
originates from the reduced distance of Rosetta from the nucleus
(Fig.~\ref{fig:copsfit}, 2nd panel), while the daily variations reflect the
distribution of gas sources on the nucleus.

\clearpage
\pagebreak

\begin{figure*}
\begin{center}
\includegraphics[width=0.7\textwidth]{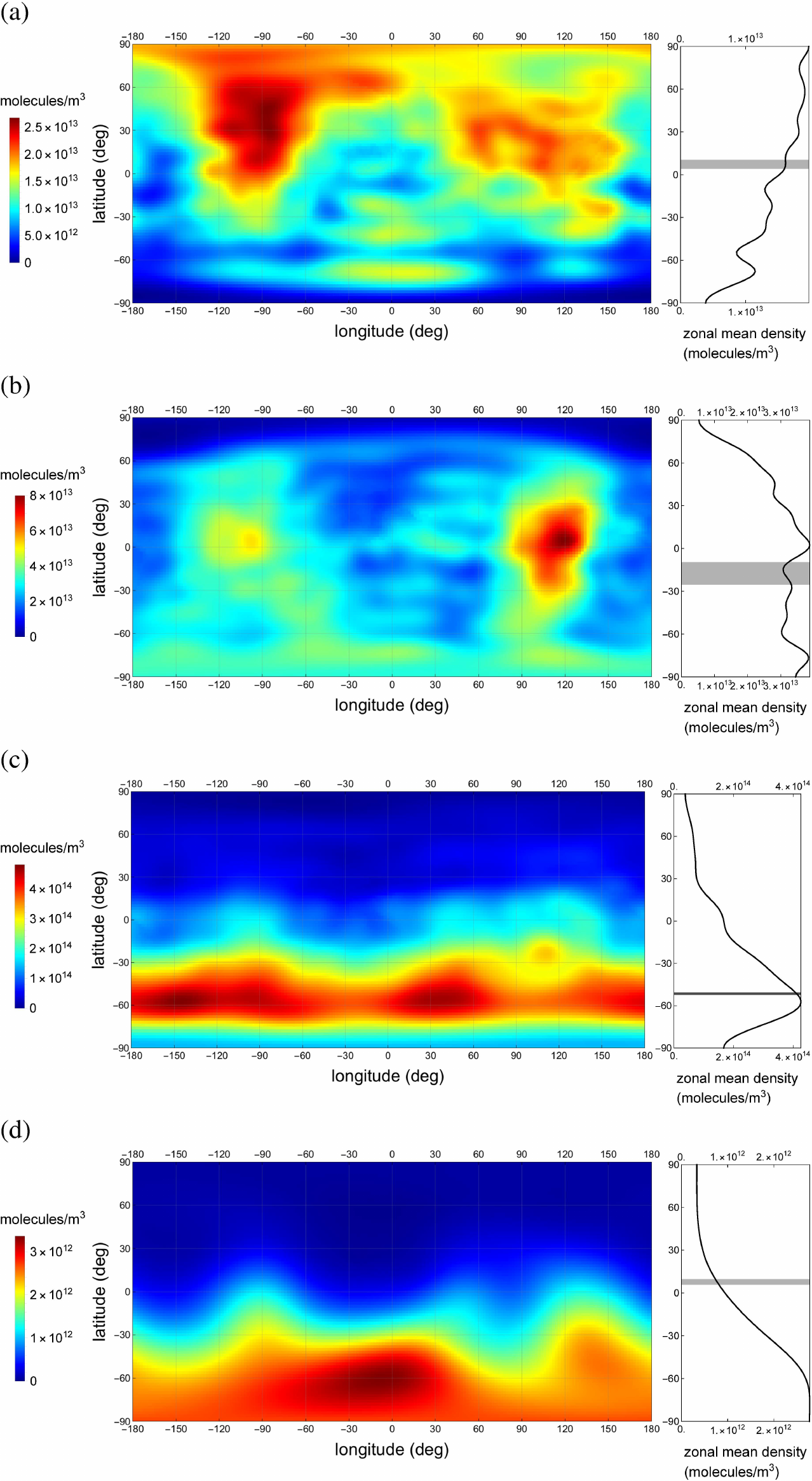}
\end{center}
\caption{
Model gas density at distance 100~km from the comet for (a) \aprilXV{}, (b) \juneXV{}, (c) \septemberXV{}, and (d) \mayXVI{}.
The side panels show the zonal mean density, the grey bars denote the subsolar latitude during the observation time.
The highest gas densities shift from the northern to the southern hemisphere.
}
\label{fig:sphere100}
\end{figure*}

\subsection{Gas density at 100~km distance and on the surface}

The COPS measurements provide a good sampling of the entire latitude/longitude range of the nucleus, but at varying radial distance (Fig.~\ref{fig:orbit}).
The parametrisation of the data set in terms of the gas model makes it possible to reconstruct the COPS data set on a spherical surface at fixed distance 100~km, see Fig.~\ref{fig:sphere100}.
The small error between the COPS data and the model during the entire monthly data set is an indication that the stationary approximation provides a suitable description of the emission process.
A continuous gas release is reflected in similar, repetitive density variations along Rosetta's orbit, which appear for similar Rosetta-nucleus viewing geometries (Fig.~\ref{fig:copshourly}).

The 100~km densities show a 3 fold increase of the absolute gas density towards perihelion from $3\cdot 10^{13}$~m$^{-3}$ (\aprilXV{}) to  
$5\cdot10^{14}$~m$^{-3}$ shortly after perihelion (\septemberXV{}).
In \mayXVI{}, the maximum density drops to $3\cdot 10^{12}$~m$^{-3}$.
In this month, the emitted gas is taken to be CO$_2$.
The seasonal effects are shown in the zonal mean densities (lateral panels in Fig.~\ref{fig:sphere100}), which undergo a systematic shift of highest densities from northern to southern latitudes.
One month before equinox (\aprilXV{}), the peak zonal density stems from two areas at (longitude,latitude)=$(-90^\circ,45^\circ)$ and $(100^\circ,20^\circ)$.
Both areas are still visible in \juneXV{} and show increased densities of $5\cdot10^{13}~\mathrm{m}^{-3}$ and 
$8\cdot10^{13}~\mathrm{m}^{-3}$, respectively.
Summer solstice (September 5, 2015) follows $23$~days after perihelion.
Around this time, the southern hemisphere received the largest solar radiation
and the zonal density in \septemberXV{} attains its highest value
at the corresponding sub-solar latitudes, see Fig.~\ref{fig:sphere100}c.
In the other months, the maxima in the zonal mean density are not occurring at the mean sub-solar latitude during the corresponding observation month, indicated by the bars in the zonal panels.
In particular, CO$_2$ sources in the southern hemisphere are still present in the zonal density distribution for
\mayXVI{}, Fig.~\ref{fig:sphere100}d, long after perihelion and after outbound equinox, which puts the sub-solar latitude to the northern hemisphere.

\clearpage
\pagebreak

\subsection{Total and surface emission rates}

\begin{figure}
\begin{center}
\includegraphics[width=0.8\columnwidth]{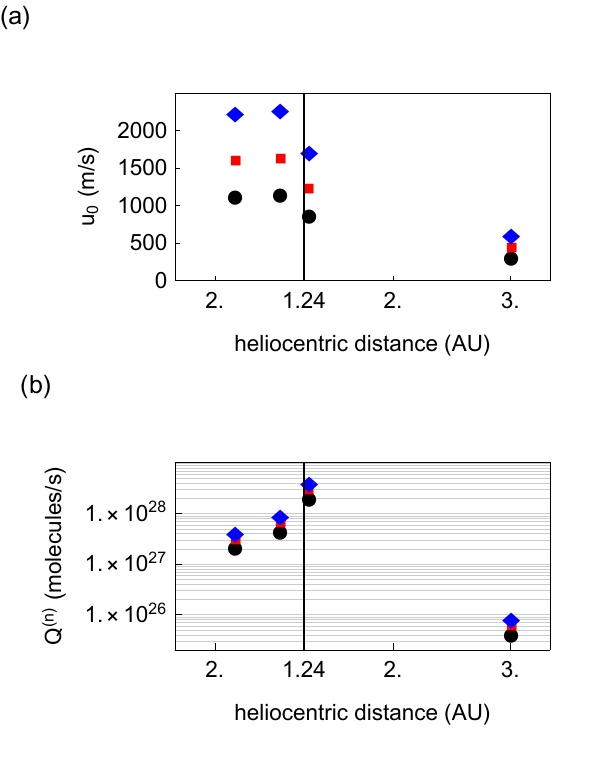}
\end{center}
\caption{
(a) Velocity $u_0$ leading to best agreement with COPS data and (b)
total volatile emission rate for (from left to right), \aprilXV{}, \septemberXV{}, \juneXV{}, and \mayXVI{}.
The symbols denote three different coma temperatures: $50$~K (circle), $100$~K (square), and $200$~K (diamond).
The perihelion distance is marked by the vertical line.
\label{fig:qtotal}}
\end{figure}

\begin{figure*}
\begin{center}
\includegraphics[width=0.7\textwidth]{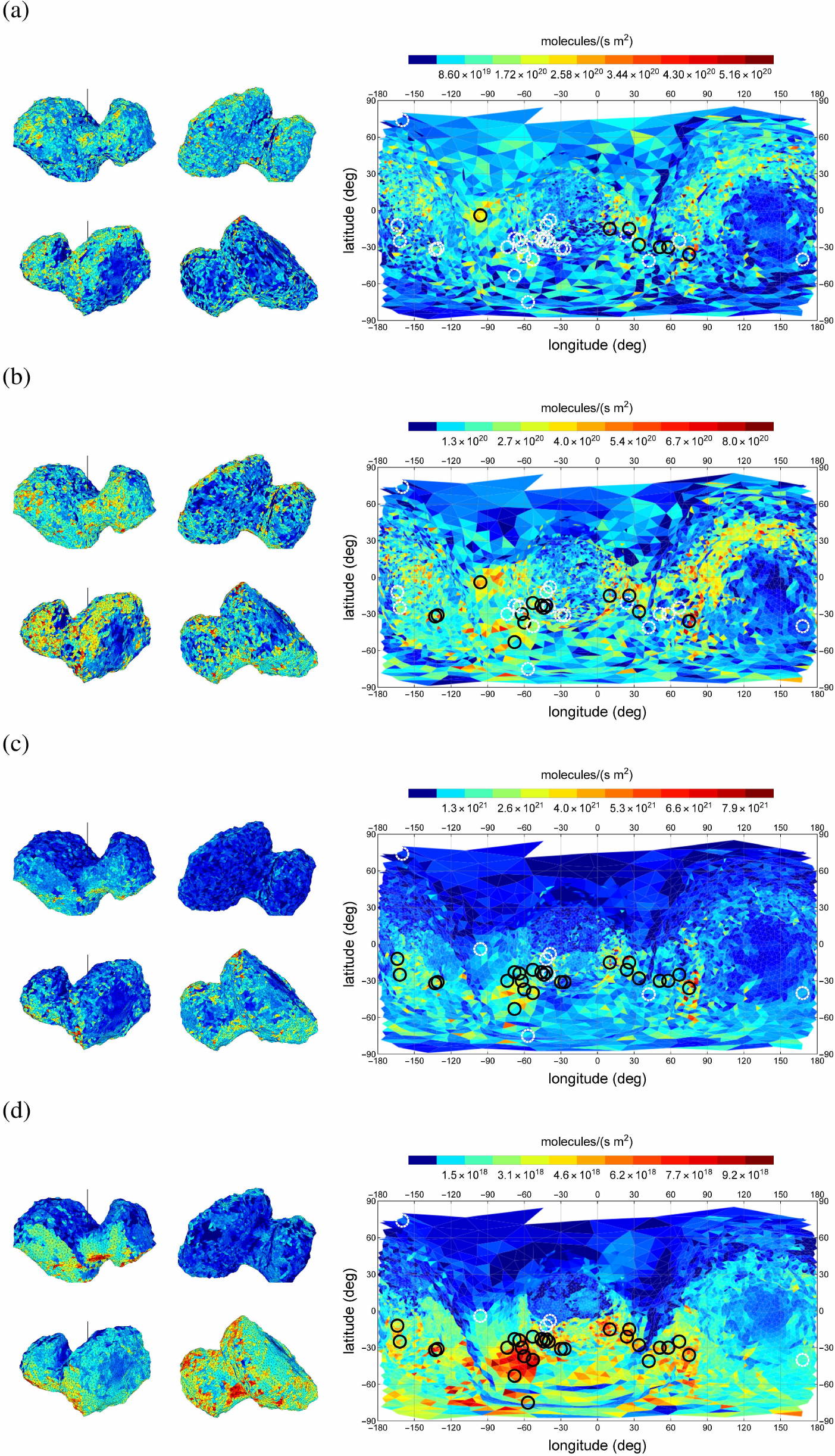}
\end{center}
\caption{\label{fig:activitycards}
Surface gas emission rate for (a) \aprilXV{}, (b) \juneXV{}, (c) \septemberXV{}, and (d)
\mayXVI{} originating from the inverse gas model for the COPS data set.
On the left, the different 3d-views include the rotational axis.
On the right, a global map is shown.
In addition to the general migration of gas sources from northern parts to the
southern parts, more localised gas source regions are revealed.
The circles indicate locations of dust outbursts during perihelion reported by \protect\cite{Vincent2016}.
Solid black circle: correlation between dust outbursts and gas emitting areas, dotted white circle: no correlation}.
\end{figure*}

\begin{figure*}
\begin{center}
\includegraphics[width=0.8\textwidth]{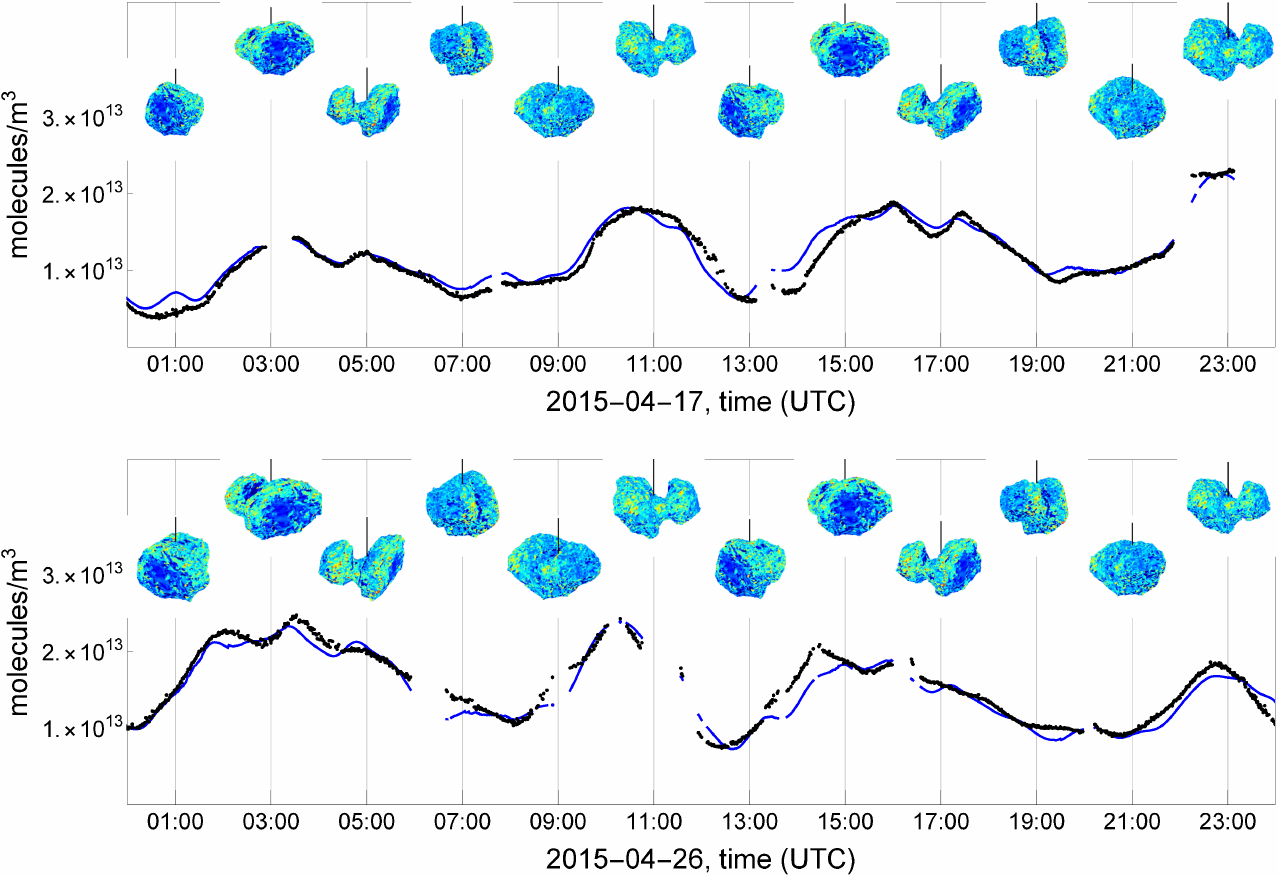}
\end{center}
\caption{
Diurnal variation of the gas density for two different days (April 17 and April 26, 2015), COPS data is marked by black dots, the blue line shows the modeled gas density.
The inset shows the momentary view from Rosetta's orbital position.
Rosetta surveys similar surface areas on both days and the gas densities show a repetitive pattern.
This indicates spatial and temporal permanence of the gas sources.
\label{fig:copshourly}}
\end{figure*}

The gas model in Sect.~\ref{sec:model}, together with the fit to the
COPS data in Sect.~\ref{ssec:inverse}, allows us to obtain the outflow
velocity $u_0$, the surface emission rates $\dot{\rho}_{i,{\rm surf}}$,
and the total production rate $Q$.
The fit to the COPS densities in Sec.~\ref{ssec:fit} yields the best
value for the parameter $U_0$, but does not set the absolute value of
the coma gas velocity $u_0$.
To obtain emission rates requires to specify the velocity field Eq.~\eqref{eq:velfield},
which includes the parameter $\beta$ and through it the coma gas
temperature $T$.
According to \cite{Gombosi1986,Biver2015,Rubin2014,Huang2016},
adiabatic cooling of the gas during expansion into vacuum leads to
a reduction in gas temperature compared to the surface temperature.
We study the parameter sensitivity by considering three
different temperatures $T=50,100,200$~K.
The resulting range of outflow velocities $u_0$ and total production
rates $Q$ is shown in Fig.~\ref{fig:qtotal}.
The total production rate increases towards perihelion and reaches
$2.0 \cdot 10^{28}$--$4.0 \cdot 10^{28}$~molecules/s shortly after perihelion (\septemberXV{}).
The values obtained here are within the ranges shown in Fig.~9 of \cite{Hansen2016}.
In \mayXVI{} the total production rate has dropped to $4.1$--$8.1 \cdot 10^{25}$~molecules/s.
Besides the total production rate, the gas model underlying the COPS fit determines the local surface emission rates with a high spatial resolution.
For the case of a coma gas temperature of $T=100$~K, the corresponding surface emission is shown in
Fig.~\ref{fig:activitycards} according to Eq.~\eqref{eq:faceprod}.
Similar surface regions are active in April and June 2015.
The two density maxima described in Fig.~\ref{fig:sphere100}a-b may be attributed to widespread gas sources around 
(longitude,latitude)=($-90^\circ$,$0^\circ$) (Anuket region) and ($60^\circ$ to $90^\circ$,$0^\circ$ to $-45^\circ$) (Aten, Khepry regions).
The overall drift of the density maxima from north to south in
Fig.~\ref{fig:sphere100} is reflected in the changes of the emission rates.
Directly after perihelion in \septemberXV{}, the water emission from the surface resembles the illumination conditions during that time (see Fig.~2 in \cite{Lai2016}).
Several months later (\mayXVI{}),
the surface emission comes from
the southern hemisphere, Fig.~\ref{fig:activitycards}d, and shows the
decoupling of the CO$_2$ gas production from the sub-solar
latitude at that time, see Tab.~\ref{ta:months}.
The same localised source regions are observed for \septemberXV{} and \mayXVI{} 
around ($-60^\circ$,$-30^\circ$ to $-45^\circ$),
at ($10^\circ$,$-15^\circ$), and at ($75^\circ$,$-40^\circ$).
After perihelion, we find a correlation of more active gas emitting areas
(flux exceeds 3.5 times the average gas flux) with the locations of previous
short (less than 5 min) dust outbursts observed optically around perihelion
\citep{Vincent2016}, marked by black circles in all surface maps.
In each of the two months \septemberXV{} and \mayXVI{}, 29 out of 34 reported
dust outburst locations are close to active gas emitters, while in \juneXV{}
only 8 and in \aprilXV{} 15 locations match.
This observational correlation might indicate a persistence of CO$_2$ sources
at dust outburst locations over 9 months.  \\
To illustrate how the surface sources are leading to specific COPS densities,
Fig.~\ref{fig:copshourly} shows two periods of 24 hours for COPS data and the
modeled density.
These periods are separated by 9 days
in \aprilXV{} and characterise similar overflight positions for Rosetta,
see Fig.~\ref{fig:copsfit} for the complete data set.
For both days, the gas densities give repetitive patterns for
succeeding relative minima and maxima.
The differences of magnitude are mostly related to slightly different
Rosetta distances and are equivalent to a $x^{-2}$ drop in the coma.
These patterns show that the model assumption of time-independent but
spatially varying surface emission rates results in a good agreement to
the temporally varying COPS data.

\clearpage
\pagebreak

\clearpage

\section{Summary}

Within this article, we have presented an inverse gas model approach
for COPS data measurements around the comet 67P/C-G.
The least-squares analysis leads to a solution for the given shape
and set of observations.
Coupled with the numerically efficient evaluation of the gas model, it reconstructs the COPS data spanning several weeks from a spatially varying surface emission map.
The total production rates are in agreement with the combined ground observation and Rosetta data \citep{Hansen2016}.
The surface maps support the hypothesis of a widespread gas release across the surface in combination with additional localised gas sources.
Localised source regions are consistently retrieved by the inverse model over distinct data sets separated by more than 9 months.
A similar model has been applied to describe the dust emission of 67P/C-G, with a similar  correlation ($>0.9$) between optical dust observations and theoretical prediction \citep{Kramer2015a,Kramer2016}.
Comparing different seasonal conditions before and after perihelion, 
we observe a systematic southward migration of the gas sources.
In particular, several months after perihelion the gas sources and the sub-solar latitude are not coinciding.
This points to a north-south dichotomy with respect to the gas release.

\section*{Acknowledgements}

The work was supported by the North-German Supercomputing Alliance (HLRN).
We thank M.~Malmer for helpful discussions.
Rosetta is an ESA mission with contributions from its member states and
NASA. We acknowledge herewith the work of the whole ESA Rosetta team. 
Work on ROSINA COPS at the University of Bern was funded by the State of Bern,
the Swiss National Science Foundation, and by the European Space Agency
PRODEX program.

\bibliographystyle{mnras}

\end{document}